# Molecular Dynamics Simulation of Vacancy Cluster Formation in β- and α-Si$_3$N$_4$

E. Adabifiroozjaei[1,*], S.S. Mofarah[2], H. Ma[2], Y. Jiang[2], M. Hussein N. Assadi[2,3], T.S. Suzuki[1]

[1]*Research Center for Functional Materials (RCFM), National Institute for Materials Science (NIMS), Tsukuba 305-0047, Japan.*
[2]*School of Materials Science and Engineering, UNSW Sydney, Sydney, NSW 2052, Australia.*
[3]*Center for Green Research on Energy and Environmental Materials (GREEN), National Institute for Materials Science (NIMS), 1-1 Namiki, Tsukuba, Ibaraki 305-0044, Japan.*

## ABSTRACT

Molecular dynamics simulation is used to study vacancy cluster formation in β- and α-Si$_3$N$_4$ with varying vacancy contents (0 - 25.6 at%). Vacancies are randomly created in supercells, which were subsequently heat-treated for 114 nanoseconds. The results show that both β and α can tolerate vacancies up to 12.8 at% and form clusters, confirming previous experimental data indicating 8 at% vacancy in α-Si$_3$N$_4$. However, 25.6 at% vacancy in β results in complete amorphization, while the same amount in α results in a transformation of a semi-amorphous α phase to a defective β phase, leading to the removal of the clusters in newly formed β. This clearly explains why cluster vacancies are not experimentally observed in β, considering that β-Si$_3$N$_4$ ceramics are produced from α. Furthermore, the lattice parameters of both modifications increase with increasing vacancy content, revealing the cause of different lattice constants that were previously reported for α-Si$_3$N$_4$.

## INTRODUCTION

Silicon nitride (Si$_3$N$_4$) possesses excellent high-temperature stability, high thermal shock resistance, good creep behavior, and good corrosion and oxidation resistance [1-3]. It is very well known that lattice defects, particularly Schottky defects, play an important role in determining the overall properties of Si$_3$N$_4$ ceramics [4, 5]. The Schottky defects are either caused by the introduction of oxygen as an impurity or thermal annealing that results in the formation of simultaneous Si and N vacancies [6]. In the case of oxygen impurity, vacancies are created to maintain the overall charge neutrality of the lattice [7]. Although α-Si$_3$N$_4$[a] initially was assumed to be established by oxygen impurity, but later research has shown that nearly pure α-Si$_3$N$_4$ can be synthesized [7]. For α-Si$_3$N$_4$, previous research works showed the presence of vacancy clusters (as large as 30 nm) or dislocation loops [8]. Such clusters or loops are not yet detected in β-Si$_3$N$_4$, seemingly due to the unstrained lattice of β-Si$_3$N$_4$ (while α-Si$_3$N$_4$ is strained) impeding Schottky defects formation. Moreover, it is assumed that Si and N mobilities are higher in β-Si$_3$N$_4$ than in α-Si$_3$N$_4$, so vacancies are assumed to get vanished in the latter during high-temperature sintering [9]. This reasoning is based on the diffusion data of Kijima and co-workers [10], which are controversial, since, in spite of all similarities of β and α modifications, the activation energies of N diffusion in the former and latter are measured as 778 and 235 kJ/mol, respectively. Wang *et al.* [11] calculated the amount of Schottky defects in α-Si$_3$N$_4$ sintered at 1973K to be 0.08 at%. This calculation is only based on a large cluster of vacancies (~ 25 nm in diameter), ignoring the smaller ones. Accordingly, using $D_N = 4D_v[V_{Si_3N_4}]$, in which $D_N$, $D_v$, and $V_{Si_3N_4}$ are diffusion coefficient of N, diffusion coefficient of vacancy (estimated as $2 \times 10^{-18}$ m$^2$/s), and volumetric vacant fraction of Si$_3$N$_4$, respectively, Wang *et al.* [11] calculated the $D_N$ in α-Si$_3$N$_4$ as $6 \times 10^{-21}$ m$^2$/s. This is stated to be close to that obtained by extrapolation of Kajima's data at 2000 K ($10^{-19}$ m$^2$/s) [10], although being two orders of magnitude different. Additionally, the extrapolation of Kajima's data seems to be considerably (three orders of magnitude) miscalculated, since extrapolation of Kajima's work yields diffusion coefficient of N as ~ $10^{-22}$ m$^2$/s at 2000 K. The recent more accurate measurement by Schmidt [12] demonstrated that $D_N$ and $D_{Si}$ in both modifications are in the same orders of magnitude and $D_N$ in α-Si$_3$N$_4$ is ~ $6 \times 10^{-19}$ m$^2$/s at ~ 2000 K. Putting this in the formulae that Wang used, and considering $D_v$ as $2 \times 10^{-18}$ m$^2$/s, the $[V_{Si_3N_4}]$ is estimated as ~ 8 at%. Considering these discrepancies, in the present study, we have used molecular dynamics simulation (MD) to investigate the tolerance of both modification of silicon nitride to contain varying content of Si and N vacancies (0 – 25.6 at%) as well as their tendency to form cluster when heat-treated for long time (114 ns).

## SIMULATION PROCEDURE

MD simulation was carried out by QuantumATK (classical forcefield) package [13], with Virtual NanoLab (VNL) for visualization. The forcefield used in this study is Marian-Gastreich two-body potential [14], which is used quite frequently for simulation of silicon nitride ceramics [15-18]. Furthermore, it was compared recently [19] with other existing forcefields for silicon nitride ceramics and was shown to be the best for reproducing properties of amorphous and crystalline systems of silicon nitride. In the present study, both perfect α (5 × 5 × 5 supercell) and β (5 × 5 × 10 supercell) systems were used as initial systems that each contained 1500 of Si and 2000 of N. These limited number of atoms are selected in order to conduct long simulation runs for each supercell (114 ns). The imperfect crystals were prepared from the initial cells by removing 0.4, 0.8, 1.6, 3.2, 6.4, 12.8, and 25.6 at% of Si and 0.4, 0.8, 1.6, 3.2, 6.4, 12.8, and 25.6 at% of N. First the perfect crystal (called 0 hereafter)[b] was heated in canonical ensemble (NPT) from 300 K to 2100 K[c] with interval of 100 K and annealing time of 3 ns at each temperature and then cooled down again to 300 K with the same path as heating (so 57 ns heating and 57 ns cooling).

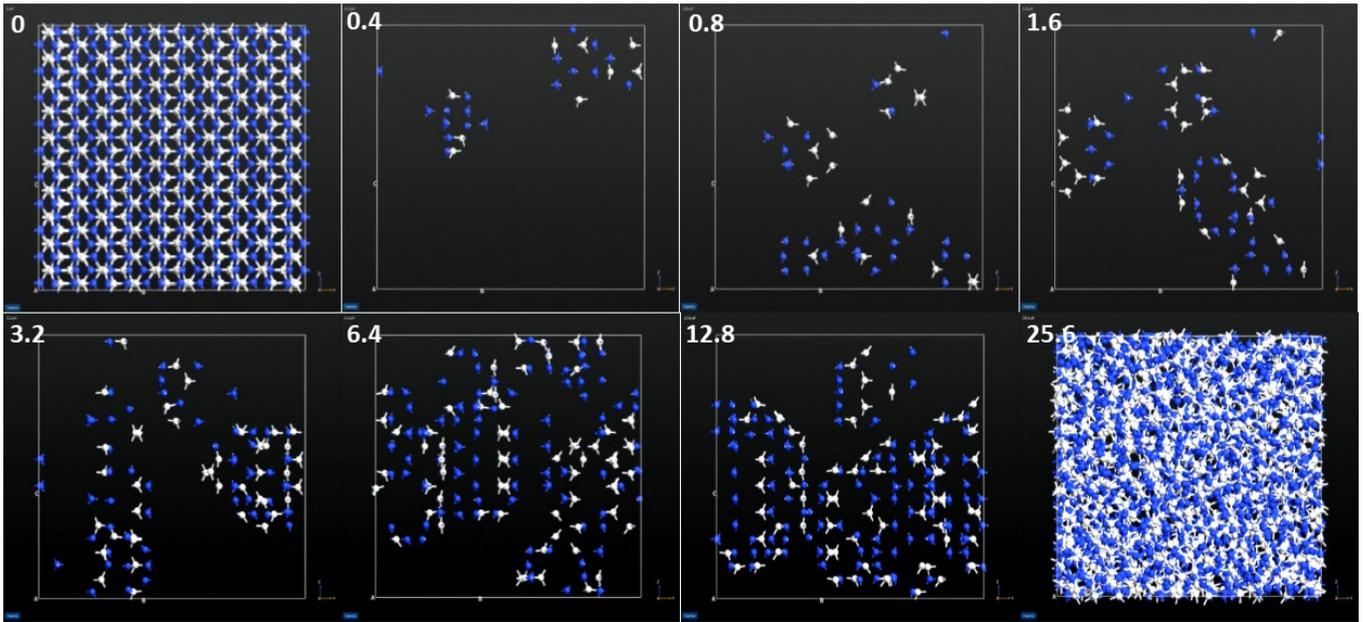

**Figure 1.** Vacancy cluster visualization in different systems of β-Si$_3$N$_4$ with various defect contents cooled to 300 K. The content of defects is printed in left-top corner of each subsection. Blue and white spheres represent N and Si atoms, respectively.

Next, 0.4 lattice was created from annealed perfect lattice by randomly deleting Si and N atoms from the perfect annealed lattice. Similar heating and cooling procedure were followed for 0.4 system. Other systems were prepared following a similar procedure; 0.8 from 0.4, 1.6 from 0.8 and so on. This procedure was employed to avoid any unwanted sudden heating or destruction in the lattices caused by relatively high defect content. Additionally, since cells with a higher content of vacancies are made from the preceding ones, higher containing vacancies cells have a longer heat-treatment history (*e.g.,* 12.8 cell is heat-treated for 798 ns, being close to real experimental time). Each step in the runs was set to one femtosecond (fs). The periodic boundary condition is applied in all directions for all cells. In systems with a high percentage of vacancies, vacancy clustering creates voids throughout a significant portion of the cell. The structural distortions caused by these voids render significant numbers of lattice sites near the voids to deviate from the local periodicity, or local coordination environment, of the hexagonal lattice. The pressure in all runs was $1 \times 10^5$ Pa, and the temperature and pressures are controlled by the bath at every 10 and 500 fs. The Nosé-Hoover Chain thermostat [20] in combination with Martyna–Tobias–Klein barostat [21] were used to maintain the temperature and pressure, respectively.

## RESULTS

**Figure 1** shows the lattice of different systems of β-Si$_3$N$_4$ with various contents of defects. Except in 0 and 25.6 systems, in the rest, for better illustration of the cluster vacancy formation, Si atoms with a coordination number of 4 and N atoms with a coordination number of 3 are deleted.[d] Therefore, the atoms left in these Figures are either under-coordinated or over-coordinated. These atoms form a crust in which vacancies are present. An interesting point is that even at very low defect content (0.4 at%), the introduced vacancies form clusters, although being small in size. With the increase in the defect content, the size of these clusters significantly increases. In 12.8 system, the size of these clusters approaches approximately 1/3 of the lattice size (10.67 Å out of 29.18 Å of the cell dimension), as indicated by the comparison of mass density profiles of 12.8 and 0 systems along the C axis shown as **Figure 2a**. The cluster size also can be approximated visually from Supplementary **Figure S1** which shows only under- or over-coordinated atoms and Supplementary **Figure S2** which shows all atoms. In spite of such large clusters of vacancies, the crystalline structure does not transform into an amorphous structure even at a temperature as high as 2100 K. In contrast, such transformation occurs in systems with a higher content of vacancy such as 25.6. Crystalline to amorphous phase transformation is expected when the number of broken bonds exceeds a certain threshold [22, 23], leading to thermodynamic instability of the crystalline phase, which is believed to occur either due to extensive chemical disordering or compositional shift from the exact stoichiometry [22]. In some instances, the amorphous phase can, nonetheless, be thermodynamically more stable because of its atoms having a higher coordination number than the atoms in the crystalline phase [23]. In the present work, as **Figure 2b** shows, the percentage of Si and N with higher coordination number is higher in 25.6 system, which is amorphous, than the 12.8 crystalline system. It is also reported that ordered vacancy can be the starting point of amorphization, particularly at temperatures close to melting temperature [24]. Moreover, nanovoids or vacancy clusters are reported to promote crystalline to amorphous phase transformations too [25]. Another important point is, as seen in **Figure 1** and Supplementary **Figure S1**, that the majority of under- or over-coordinated atoms that shape the crust of vacancy cluster are ordered. This along with the fact that all of the randomly-created vacancies before heat-treatment were disordered and distant from one another highlight the high tendency of Si and N for forming ordered defects in β-Si$_3$N$_4$. Defect clustering in silicon nitride ceramics is believed to be a very difficult process due to the dominance of directional covalent bonds which impedes atoms mobility—a requirement for such large vacancy cluster formation. However, the very high bond energy of Si-N (439 kJ/mol [26]) has to be also considered. A high energy penalty associated with bond breakage can force the system to have the lowest number of broken bonds, which can be achieved by clustering. These data clearly show that, if any Schottky defects are created in β-$Si_3N_4$, either due to thermal heating or oxygen impurities, then they form cluster vacancies. However, in next section it will be shown that why in experiment cluster formation has not yet been observed in β-$Si_3N_4$.

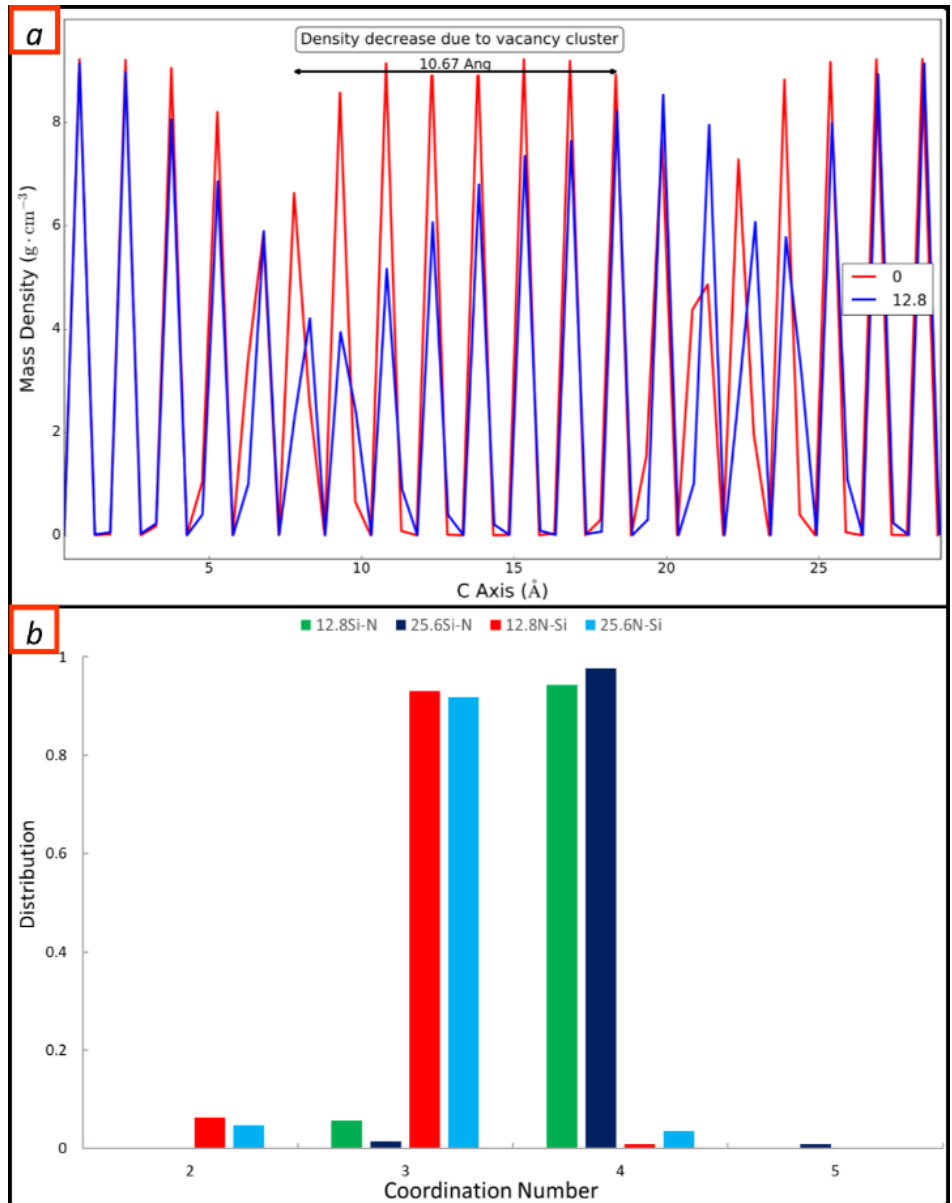

**Figure 2.** a) Comparison of mass density profiles of 0 and 12.8 systems indicating density decrease due to vacancy cluster formation, b) Comparison of coordination number of N and Si atoms in crystalline (12.8) and amorphous system (25.6).

**Figure 3** shows the structure of the various systems of α-$Si_3N_4$ after the annealing process. Similar to β-$Si_3N_4$ systems, the vacancy cluster form in the imperfect systems. However, the sizes of clusters are smaller and more uniformly distributed across the lattice. Additionally, less ordering of under- or over-coordinated atoms are observed, when compared with those of β-$Si_3N_4$. Supplementary **Figure S3** (only under- or over-coordinated atoms) and Supplementary **Figure S4** (all atoms included) show rotated views of the extended defects in 12.6 system of α-$Si_3N_4$, which can be compared with that of β-$Si_3N_4$ given in Supplementary **Figure S1** and **Figure S2**. Also, in contrast to 25.6 system of β-$Si_3N_4$, complete amorphization does not occur in α-$Si_3N_4$. Instead, as seen in Supplementary **Figure S5**, a semi-amorphous phase of α transforms into a mixture of α and β at 1900 K. The simulated X-ray pattern of the average structure before (at 1700 K and 1800 K) and after phase transformation (2000 K) is shown in **Figure 4a**. As seen, peaks belonging to both α and β phases are present in the simulated pattern, indicating that the complete transformation to β is not reachable in nanoscale-time simulations conducted in the present study. This was confirmed through more 20 ns heat-treatment of the system at 2000 K (**Figure 4b** and **Figure 4c**) showing the formation of higher order in the lattice, indicating that time plays a critical role in the completion of the transformation.

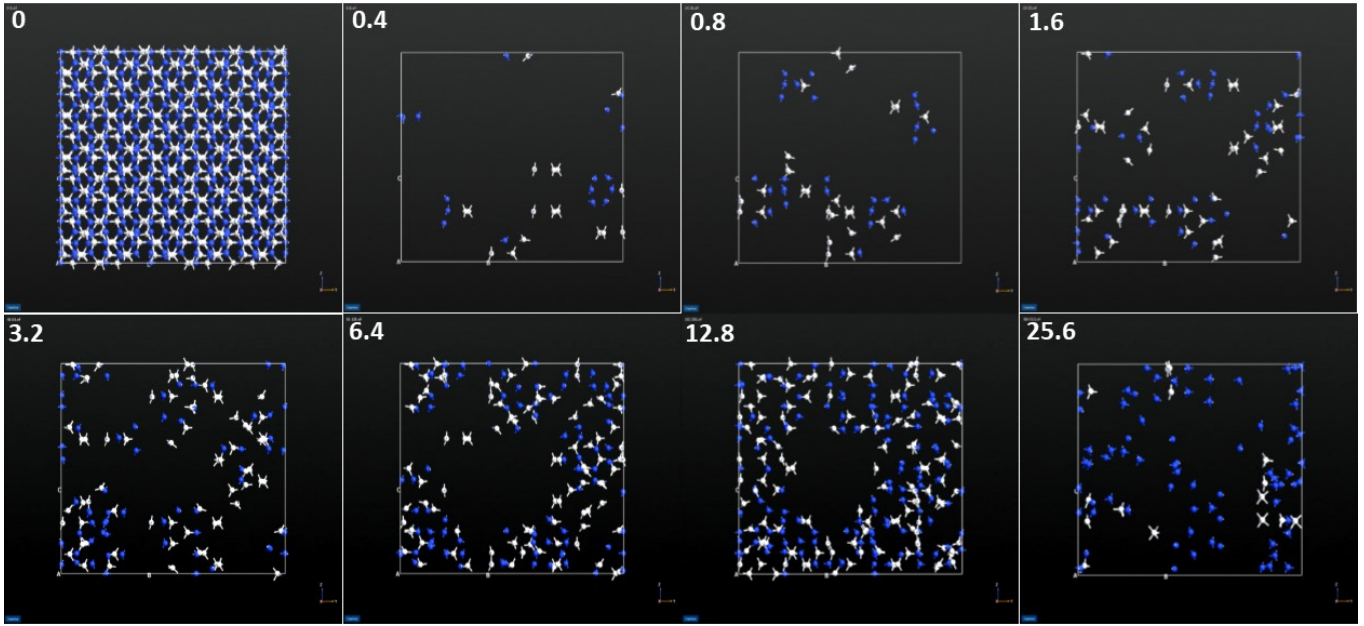

**Figure 3.** Vacancy cluster visualization in different systems of α-Si$_3$N$_4$ with various defect contents cooled to 300 K. The defect content is printed in the top-left corner of each subsection. Blue and white spheres represent N and Si atoms, respectively.

This is particularly applicable to silicon nitride because of the very slow diffusion rates of Si and N in the lattice [10, 12]. Besides the peaks of α and β phases, there are also peaks in the X-ray pattern of transformed samples that cannot be assigned to a specific phase. These peaks might be related to planes that are formed temporarily during the phase transformation, or from the defects that form planes at moderate distances relative to original planes of silicon nitride lattices. Between 300 K and 1900 K, the 25.6 system of α exists as a semi-crystallized phase, although with an increase in the temperature, the previously formed cluster of vacancies gradually break down, but never completely get vanished. With the phase transformation occurring, all of the vacancy clusters disappear, although still some under- or over-coordinated atoms (mostly N) are left in the structure (**Figure 3**). These atoms do not form any vacancy cluster and mostly belong to the portions of the lattice that are not still completely transformed. The disappearance of the vacancy cluster during phase transformation is most likely due to the high mobility of the atoms. Since β-Si$_3$N$_4$ sintered samples are usually obtained by heat-treatment of α powders, therefore the disappearance of vacancy cluster during α to β phase transformation could be used as strong evidence for why vacancy clusters or dislocation loops are not observed in β samples obtained from sintering of α powders. However, this does not rule out the possibility of the clustering of defects in β as long dislocations, which have been observed in experiments [27, 28]. Additionally, it has to be noted that β samples prepared by other methods [29] or neutron-irradiated [30] can have vacancy clusters as also observed in the present work (**Figure 1**, 12.8 system). According to a previous study, in the absence of liquid phase, α to β phase transformation occurs at a temperature above 1873 K [8], being very close to the transformation temperature observed in the current simulation (1900 K). However, it is also reported that single crystals of α only transform to β after heat-treatment at 2423 K [31]. This combined with current simulation observations highlight the effect of the defects in the enhancement of phase transformation in silicon nitride.

**Figure 4d** and **Figure 4e** show the change in the lattice parameter of β and α systems with the change in vacancy content, respectively. Since both *a* and *c* exactly followed the same trend in both systems, only *a* is shown and discussed here. Accordingly, with an increase in vacancy content in both β and α, the lattice parameter generally increases. The increase in β is more enhanced, although in 12.8 system, the lattice parameter slightly dips from the value of the less defective 6.4 system. In α-Si$_3$N$_4$, *a* increases abruptly for 0.4 system relative to 0 system, while for 0.4 β system such abrupt increase is not observed. This trend might be unexpected since β system contains a larger vacancy cluster than α (**Figure 1** and **Figure 3**), which has more uniform defect distribution with smaller lattice parameters. To explain the origin of this peculiarity, the lattice constants and the volume of 0.4 and 0.8 systems of α and β systems after 3 ns heat treatment at 300 K are compared with those of the same systems heat-treated for 114 ns (from 300 K to 2100 K, and cooled down to 300 K). The results are given in **Table 1**. The main difference here is that in systems heat-treated for 3 ns, there was no enough time for vacancy clustering, while in those heat-treated for 114 ns, the chance of vacancy clustering was quite high. The results in **Table 1** clearly indicate that systems with more individual vacancy defects (uniformly distributed defects; pale blue columns) have higher volume and lattice constants than those with clustered defects (large cluster vacancy; pale orange columns). Therefore, it can be concluded that the larger the cluster size is, the smaller volume and lattice constants are. It can also be inferred that, in addition to the decrease in the energy of the systems, the decrease in volume (or equivalently the increase in density) might also explain why vacancies tend to form large clusters in silicon carbide polymorphs. After the abrupt increase, the parameter increases with a moderate slope till 12.8 system. This change in lattice parameters of α-Si$_3$N$_4$ is also observed in experiments, and it is mainly attributed to the presence of vacancy clusters and dislocation loops [6].

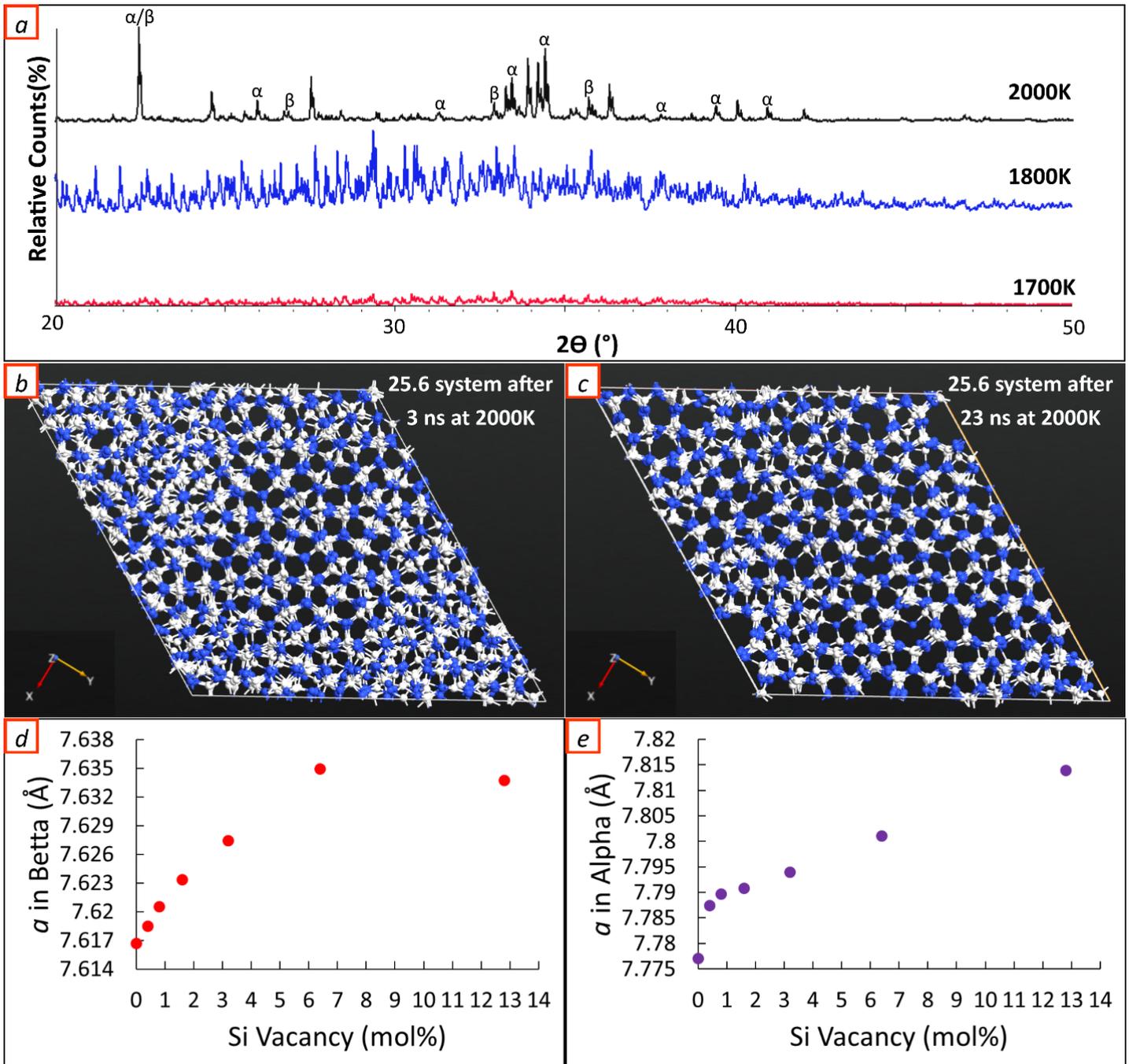

**Figure 4.** a) X-ray pattern of 12.8 system of α-Si$_3$N$_4$ at 1700 K, 1800 K, and 2000 K, b and c) comparison of 25.5 system after 3 ns and 23 ns heat-treatment at 2000 K, respectively, d and e) change in lattice parameter (*a*) of β and α systems *vs.* change in vacancy content, respectively.

**Table 1.** Comparison of volume and lattice constants of 0.4 and 0.8 systems of α and β heat-treated for 3 ns (pale blue columns) and 114 ns (pale orange columns).

|  | 0.4 α 3ns | 0.4 α 114ns | 0.4 β 3ns | 0.4 β 114ns | 0.8 α 3ns | 0.8 α 114ns | 0.8 β 3ns | 0.8 β 114ns |
|---|---|---|---|---|---|---|---|---|
| **V (Å$^3$)** | 296.4185 | 296.4096 | 146.4316 | 146.4228 | 296.6768 | 296.6696 | 146.6662 | 146.5404 |
| ***a* (Å)** | 7.7872 | 7.7869 | 7.6196 | 7.6185 | 7.7898 | 7.7897 | 7.6227 | 7.6205 |
| ***c* (Å)** | 5.6436 | 5.6435 | 2.9135 | 2.9129 | 5.6454 | 5.6454 | 2.9146 | 2.9137 |

This is confirmed by current results showing that cluster vacancy formation can cause variation in the lattice constants of silicon nitride. In fact, the variation of the lattice constant ($a$) observed in the current study (7.777 Å to 7.813 Å) matches closely the reported experimentally (7.750 Å to 7.818 Å) [6]. Similar variations are not yet mentioned in the literature for β modification [1], since as demonstrated in the previous section, α to β phase transformation results in vanishment of vacancy clusters and corresponding effects in β modification.

## CONCLUSIONS

The current results clearly indicate that both α and β are capable of containing a considerable content of Schottky defects (up to 12.8 at%) without lattice destruction, confirming the recalculated data of Wang and co-workers (~ 8 at%). In both structures, these vacancies form considerably large-size vacancy clusters, although in β the cluster sizes are larger than in α. Introduction of more vacancies (25.6 at%) causes lattice destruction and complete amorphization in β, while in α such vacancy content results in partial amorphization at low temperature and subsequent α to β phase transformation at 1900 K, being close to that observed (1873K) in experiments for pure α-$Si_3N_4$. The vanishment of vacancy clusters during α to β phase transformation is found to be the reason why vacancy clusters are not experimentally observed in β samples. In accordance with previous experimental observation, vacancy clusters formation leads to variation in the lattice constant of α-$Si_3N_4$.

## ACKNOWLEDGMENTS

E.A. acknowledges the financial support (JSPS KAKENHI Grant Number: 18F18064) provided by the Japan Society for the Promotion of Science (JSPS).

## CONFLICT OF INTEREST

The authors declare no competing interests.

## DATA AVAILABILITY STATEMENT

All relevant data are available upon request from corresponding authors (Adabifiroozjaei.e@nims.jp.go).

## SUPPLEMENTARY INFORMATION

Supplementary information is available at https://www.sciencedirect.com/science/article/abs/pii/S0927025620301233?via%3Dihub

## REFERENCES

[1] F. L. Riley, *Silicon Nitride and Related Materials,* J. Am. Ceram. Soc. **83,** 245–65 (2000). https://doi.org/10.1111/j.1151-2916.2000.tb01182.x

[2] E. Adabifiroozjaei, P. Koshy, F. Emadi, S. S. Mofarah, H. Ma, E. Rastkerdar, S. Lim, R. F. Webster, D. R. G. Mitchell and C. C. Sorrell, *Ionic interdiffusion as interaction mechanism between Al and $Si_3N_4$,* J. Am. Ceram. Soc. **102,** 4835–47 (2019). https://doi.org/10.1111/jace.16358

[3] E. Sánchez-González, P. Miranda, F. Guiberteau and A. Pajares, *Effect of temperature on the pre-creep mechanical properties of silicon nitride,* J. Eur. Ceram. Soc. **29,** 2635–41 (2009). https://doi.org/10.1016/j.jeurceramsoc.2009.03.011

[4] J. J. Meléndez-Martínez and A. Domínguez-Rodríguez, *Creep of silicon nitride,* Prog. Mater. Sci. **49,** 19-107 (2004). https://doi.org/10.1016/S0079-6425(03)00020-3

[5] H. Yokota, H. Abe and M. Ibukiyama, *Effect of lattice defects on the thermal conductivity of β-$Si_3N_4$,* J. Eur. Ceram. Soc. **23,** 1751–9 (2003). https://doi.org/10.1016/S0955-2219(02)00374-6

[6] C.-M. Wang, X.-Q. Pan and M. Rühle, *Origin of dislocation loops in α-silicon nitride,* J. Mater. Res. **11,** 1725–32 (1996). https://doi.org/10.1557/JMR.1996.0216

[7] C.-M. Wang, *Fine Structural Features in α-Silicon Nitride Powder Particles and Their Implications,* J. Am. Ceram. Soc. **78,** 3393–6 (1995). https://doi.org/10.1111/j.1151-2916.1995.tb07984.x

[8] C. M. Wang, X. Pan, M. Rühle, F. L. Riley and M. Mitomo, *Silicon nitride crystal structure and observations of lattice defects,* J. Mater. Sci. **31,** 5281–98 (1996). https://doi.org/10.1007/BF01159294

[9] C.-M. Wang, F. L. Riley, F. Castro and I. Iturriza, *Dislocation Loops in α-Silicon Nitride,* J. Am. Ceram. Soc. **76,** 2136–8 (1993). https://doi.org/10.1111/j.1151-2916.1993.tb08350.x

[10] K. Kijima and S. I. Shirasaki, *Nitrogen self-diffusion in silicon nitride,* J. Chem. Phys. **65,** 2668–71 (1976). https://doi.org/10.1063/1.433464

[11] C.-M. Wang and F. L. Riley, *Structural nano-defects in α-silicon nitride,* J. Eur. Ceram. Soc. **16,** 679-83 (1996). https://doi.org/10.1016/0955-2219(95)00177-8

[12] H. Schmidt, U. Geckle and M. Bruns, *Simultaneous diffusion of Si and N in silicon nitride,* Phys. Rev. B **74,** 045203 (2006). https://doi.org/10.1103/PhysRevB.74.045203

[13] J. Schneider, J. Hamaekers, S. T. Chill, S. Smidstrup, J. Bulin, R. Thesen, A. Blom and K. Stokbro, *ATK-ForceField: a new generation molecular dynamics software package,* Model. Simul. Mater. Sci. Eng. **25,** 085007 (2017). https://doi.org/10.1088/1361-651x/aa8ff0

[14] C. M. Marian, M. Gastreich and J. D. Gale, *Empirical two-body potential for solid silicon nitride, boron nitride, and borosilazane modifications,* Phys. Rev. B **62,** 3117–24 (2000). https://doi.org/10.1103/PhysRevB.62.3117

[15] A. Hannemann, J. C. Schön, M. Jansen, H. Putz and T. Lengauer, *Modeling amorphous $Si_3B_3N_7$: Structure and elastic properties,* Phys. Rev. B **70,** 144201 (2004). https://doi.org/10.1103/PhysRevB.70.144201

[16] A. Hannemann, J. C. Schön and M. Jansen, *Modeling the sol–gel synthesis route of amorphous $Si_3B_3N_7$,* J. Mater. Chem. **15,** 1167–78 (2005). https://doi.org/10.1039/B415188G

[17] A. Hannemann, J. C. Schön and M. Jansen, *Stability of nanovoids in amorphous $Si_3B_3N_7$,* Phil. Mag. **85,** 2621–39 (2005). https://doi.org/10.1080/14786430500154075

[18] V.-V. Le, T.-T. Nguyen, S.-K. Kim and K.-H. Pham, *Effect of the Si content on the structure, mechanical and tribological properties of CrN/AlSiN thin films,* Surf. Coat. Tech. **218,** 87–92 (2013). https://doi.org/10.1016/j.surfcoat.2012.12.031

[19] A. Dasmahapatra and P. Kroll, *Modeling amorphous silicon nitride: A comparative study of empirical potentials,* Comput. Mater. Sci. **148,** 165–75 (2018). https://doi.org/10.1016/j.commatsci.2017.12.008

[20] G. J. Martyna, M. L. Klein and M. Tuckerman, *Nosé–Hoover chains: The canonical ensemble via continuous dynamics,* J. Chem. Phys. **97,** 2635–43 (1992). https://doi.org/10.1063/1.463940

[21] G. J. Martyna, D. J. Tobias and M. L. Klein, *Constant pressure molecular dynamics algorithms,* J. Chem. Phys. **101,** 4177–89 (1994). https://doi.org/10.1063/1.467468

[22] W. S. Lai and B. X. Liu, *Lattice stability of some Ni-Ti alloy phases versus their chemical composition and disordering,* J. Phys. Condens. Matter **12,** L53–L60 (2000). https://doi.org/10.1088/0953-8984/12/5/101
[23] N. Gao and W. S. Lai, *Relative stability of amorphous phase versus solid solution in the Ni–Ti system revealed by molecular dynamics simulations,* J. Phys. Condens. Matter **19,** 046213 (2007). https://doi.org/10.1088/0953-8984/19/4/046213
[24] B. Zhang, X.-P. Wang, Z.-J. Shen, X.-B. Li, C.-S. Wang, Y.-J. Chen, J.-X. Li, J.-X. Zhang, Z. Zhang, S.-B. Zhang and X.-D. Han, *Vacancy Structures and Melting Behavior in Rock-Salt GeSbTe,* Sci. Rep. **6,** 25453 (2016). https://doi.org/10.1038/srep25453
[25] Y. L. Li, J. Cai and D. Mo, *Molecular dynamics simulations on the effect of nanovoid on shock-induced phase transition in uranium nitride,* Phys. Lett. A **383,** 458–63 (2019). https://doi.org/10.1016/j.physleta.2018.11.017
[26] J. A. Dean and N. A. Lange 1992 *Lange's Handbook of Chemistry*: McGraw-Hill),
[27] E. Butler, *Observations of dislocations in β-silicon nitride,* Phil. Mag. **24,** 829–34 (1971). https://doi.org/10.1080/14786437108217052
[28] A. G. Evans and J. V. Sharp, *Microstructural studies on silicon nitride,* J. Mater. Sci. **6,** 1292–302 (1971). https://doi.org/10.1007/bf00552043
[29] C. Bagci, Q. Yang and W. M. Kriven, *Formation of α/β-$Si_3N_4$ nanoparticles by carbothermal reduction and nitridation of geopolymers,* J. Am. Ceram. Soc. **102,** 6542–51 (2019). https://doi.org/10.1111/jace.16545
[30] R. A. Youngman and T. E. Mitchell, *Defect aggregates in neutron-irradiated β-$Si_3N_4$,* Radiat. Eff. **74,** 267–78 (1983). https://doi.org/10.1080/00337578308218420
[31] H. Suematsu, M. Mitomo, T. E. Mitchell, J. J. Petrovic, O. Fukunaga and N. Ohashi, *The α–β Transformation in Silicon Nitride Single Crystals,* J. Am. Ceram. Soc. **80,** 615–20 (1997). https://doi.org/10.1111/j.1151-2916.1997.tb02876.x

---

[a] Silicon nitride has two major polymorphs: α with trigonal structure (space group: P$3_1c$) and β with hexagonal structure (space group: P$6_3/m$). The former is stable at temperature below 1573 K and above this it transforms to β [1].

[b] The system in the present works are named based on their defect content.

[c] Since decomposition of silicon nitride occurs at temperatures beyond 2273 K [1], 2100 K is used as maximum heat-treatment temperature to avoid any unwanted effect of decomposition.

[d] The coordination numbers of Si and N in perfect silicon nitride lattice are 4 and 3, respectively.